\documentclass[two column,prl,groupaddress,showpacs,floatfix]{revtex4}

\usepackage{graphicx}
\usepackage{dcolumn}
\usepackage{bm}
\usepackage{hyperref}
\usepackage{setspace}

\begin{document}


\title{Magneto-transport Effects in Topological Insulator Bi$_2$Se$_3$ Nanoribbons}


\author{Hao Tang}
\author{Dong Liang}
\author{Richard L.J. Qiu}
\author{Xuan P.A Gao}
\email[]{xuan.gao@case.edu}
\affiliation{Department of Physics, Case Western Reserve University, Cleveland, Ohio 44106}



\begin{abstract}
Magneto-resistance (MR) of Bi$_2$Se$_3$ nanoribbons is studied
over a broad range of temperature ($T$=300K-2K) and under various
magnetic field ($B$) orientations. The MR is strongly anisotropic
with the perpendicular MR much larger than the longitudinal and
transverse MRs. The perpendicular MR exhibits quadratic
$B$-dependence in low fields and becomes linear at high $B$.
However, when $T$ increases, the perpendicular MR becomes linear
over the whole magnetic field range (0-9T) up to room temperature.
This unusual linear MR is discussed in the context of the quantum
linear MR of the topological surface-states. We also observe the
boundary-scattering effect in MR at low temperatures, which
indicates that the out-of-plane Fermi momentum is much smaller the
in-plane Fermi momentum, excluding the simple three-dimensional
Fermi surface picture.
\end{abstract}


\maketitle

Topological insulators are a class of quantum materials that have
insulating energy gaps in the bulk, and gapless surface states on
the sample boundary that are protected by time-reversal
symmetry\cite{1,2,3}. Recently, Bi$_2$Se$_3$ and related
materials\cite{4} have been proposed as three-dimensional (3D)
topological insulators with a single Dirac cone for the surface
states. Among these materials, Bi$_2$Se$_3$, which is a pure
compound rather than an alloy like Bi$_{x}$Sb$_{1-x}$\cite{5},
owns a larger gap (0.3 eV), and is thought to be promising for
room temperature applications. The existence of a Z2 topological
phase with a surface Berry's phase in the stoichiometric compound
Bi$_2$Se$_3$ has been observed by angle-resolved photoemission
spectroscopy\cite{6,7}. To enhance the contribution of the surface
states in transport measurements, Bi$_2$Se$_3$ nanowires and
nanoribbons offer an attractive alternative to bulk samples for
studying the Dirac electrons on surface due to their high
surface-to-volume ratio\cite{8,9}. Indeed, Aharonov-Bohm (AB)
oscillations in the longitudinal MR of Bi$_2$Se$_3$ nanoribbons
were discovered, proving the existence of a coherent surface
conducting channel\cite{8}. Here, we study the MR of Bi$_2$Se$_3$
nanoribbons under various magnetic field orientations to elucidate
the transport mechanism in these novel nanomaterials. Our
measurements reveal boundary scattering effects and a linear MR in
the perpendicular field configuration that persists to room
temperature. This striking linear MR is discussed in the context
of quantum linear MR (QLMR) of systems with linear dispersion
spectrum \cite{11}, consistent with the transport through
topological surface-states.

Pure Bi$_2$Se$_3$ nanoribbons are synthesized in a horizontal tube
furnace via the vapor-liquid-solid mechanism with gold particles
as catalysts, similar to literature\cite{8,9}. Typical
Bi$_2$Se$_3$ nanoribbons have thickness ranging from 50-400 nm and
widths ranging from 200 nm to several $\mu$ms. Energy dispersive
X-ray spectroscopy analyses reveal uniform chemical composition
with a Bi/Se atomic ratio about 2$:$3, indicating the
stoichiometric Bi$_2$Se$_3$. High-resolution TEM imaging and 2D
Fourier transformed electron diffraction measurements demonstrate
that the samples are single-crystalline rhombohedral phase and
grow along the [11$\overline{2}$0] direction. The upper and lower
surfaces are (0001) planes. The as-grown samples are suspended in
ethanol by sonication and dispersed on a heavily doped Si
substrate with 300nm SiO$_2$ on its surface. Photolithography is
used to pattern four electrodes contacting single nanoribbon. The
electrodes consist of 150nm Pd with a 5nm Ti adhesion layer formed
via e-beam evaporation and lift-off. Ohmic contacts are obtained
without annealing. The transport measurements are performed in a
Quantum Design PPMS with low frequency lockin technique.

Temperature dependent four-terminal resistance $R$ of a nanoribbon
(sample $\#$1) from room temperature down to 2K is shown in Fig.
1a. Four-terminal resistance of the nanoribbons is obtained by
flowing a current $I$ (typically 0.1-1 $\mu$A) through the two
outer contacts and monitoring the voltage drop $V$ between the two
inner contacts (typical spacing ~2$\mu$m) as shown in SEM image in
Fig.1a inset. The resistance decreases with the temperature,
starts to saturate around 25K, and then remains nearly flat down
to 2K, the minimum temperature reached during these measurements.
This metallic $R(T)$ is typical behavior for heavily doped
semiconductors, and can result from the small band gap of
Bi$_2$Se$_3$ and the residual doping from intrinsic defects such
as Se vacancies\cite{carriertype}.
\begin{figure*}
\includegraphics[width=160mm]{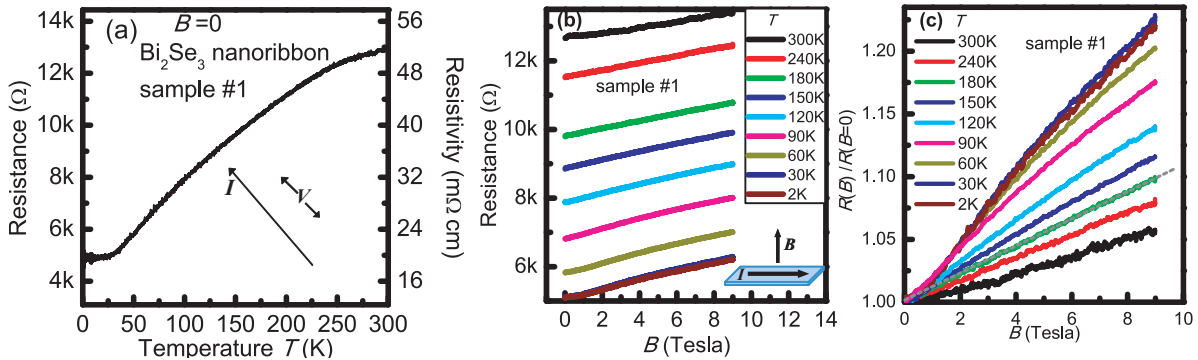}
\caption{\label{Fig.1}(color online) (a) Resistance $R$ vs. $T$ of
a Bi$_2$Se$_3$ nanoribbon (sample $\#$1). The right axis shows
resistivity. The inset shows the SEM image of device and
illustration of the four-terminal measurement of $R$. White scale
bar is 2$\mu$m. (b) $R$ vs. perpendicular magnetic field for
sample $\#$1 at various $T$ from 300K to 2K. (c) Data in (b)
plotted as $R(B)$/$R$($B$=0) vs. magnetic field. Above 90K, $R$
varies linearly with $B$. A grey dashed line is included to guide
the eye.}
\end{figure*}

The perpendicular ($B$ perpendicular to both current flow and
nanoribbon surface) MR of sample $\#$1 is shown in Fig.1b from
$T$=300K down to 2K. First of all, it is striking that there is
significant (5-25$\%$) MR for this configuration at temperatures
all the way up to 300K, and the MR curves appear to be parallel to
each other. Second, the MR has a linear $B$ dependence up to 9T,
the highest $B$ achieved in our PPMS, for temperatures higher than
90K. As $T$ decreases below 90K, $R(B)$ takes an approximately
quadratic dependence below 3T. This evolution of MR can be seen
more clearly in Fig.1c  where  the normalized resistance
$R(B)/R(B=0)$ is plotted against $B$. It has been known for a long
time that for metals with open Fermi surfaces (e.g. Au), the MR
could be linear and non-saturating at high fields\cite{10}. This
is not the case here. The existence of a linear MR for small
bandgap semiconductor could have a quantum\cite{11, 12} or
classical origin\cite{12, 13}. To explain the linear MR down to
very low fields in silver chalcogenides\cite{14}, Abrikosov first
proposed a model based on the QLMR\cite{16}for systems with
gapless linear dispersion spectrum\cite{11}. It is believed that
such gapless linear dispersion may apply for silver chalcogenide
or other small bandgap semiconductors with strong
inhomogeneity\cite{11, 12}. Without invoking the linear dispersion
spectrum, Parish and Littlewood suggested a classical origin for
linear MR in which the MR is a consequence of mobility
fluctuations in a strongly inhomogeneous system\cite{13}. For our
Bi$_2$Se$_3$ nanoribbons, the single crystal quality and small
length scale of the device rule out the models where strong
physical inhomogeneity of sample is required. Therefore, it is
tempting to attribute the linear MR observed here to the QLMR from
Dirac electrons on the Bi$_2$Se$_3$ surface, which have very small
effective mass and large cyclotron energy\cite{4}. We also point
out that Abrikosov's QLMR model predicts a $temperature$
$independent$ $\Delta$$R(B)$=$R(B)$-$R$($B$=0)\cite{11}, which is
indeed in agreement with our data in Fig.1b. Therefore, the
persistence of linear MR at high $T$ for Bi$_2$Se$_3$ nanoribbons
is striking: it could indicate the persistence of topological
surface-states induced MR up to room temperature.
\begin{figure}
\includegraphics[width=70mm]{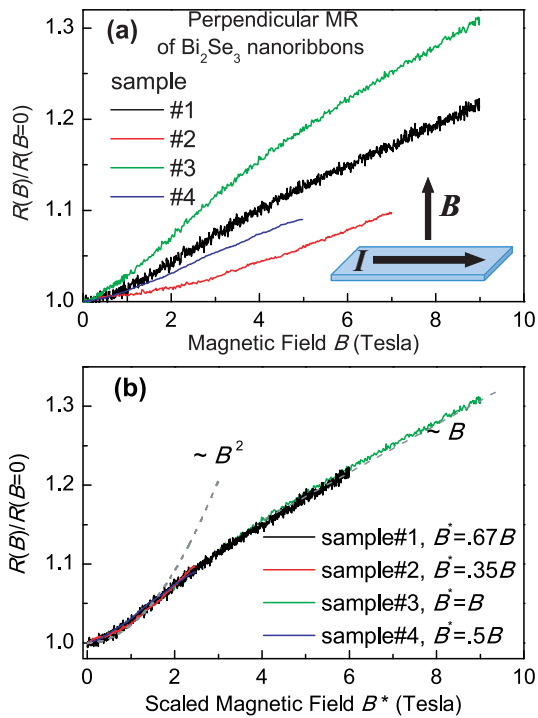}
\caption{\label{Fig.2}(color online) (a) The normalized resistance
$R(B)$/$R$($B$=0) vs. perpendicular $B$ for four Bi$_2$Se$_3$
nanoribbons. Except for sample $\#$2 where $T$=10K, all the data
are collected at 2K.  (b) $R(B)$/$R$($B$=0) vs. scaled magnetic
field $B^*$ for samples in (a). The magnetic field is scaled
linearly for sample $\#$1, 2 and 4 against $\#$3 which shows the
largest MR response. Consistent with the Kohler's rule, all MR
curves collapse onto a single curve after scaling. MR has a $B^2$
dependence at low field and linear $B$ dependence at high field as
shown by the grey dashed lines. }
\end{figure}

In an attempt to distinguish the surface electrons from bulk
electrons in our Bi$_2$Se$_3$ nanoribbons, we performed MR
measurements on several samples under various magnetic field
orientations. Fig.2a presents the normalized resistance,
$R(B)$/$R$($B$=0), as a function of the perpendicular magnetic
field for four nanoribbons. All the data were collected at $T$=2K
except for sample $\#2$ which was measured at 10K. Due to the
different carrier density and mobility of these samples, the rate
of resistance increase is different.  However, we found that all
the MR curves collapse onto a single curve if we perform a linear
scaling of the magnetic field, as shown in Fig.2b. In Fig.2b, the
magnetic field is scaled linearly for sample $\#$1, 2 and 4
against sample $\#$3 which shows the largest MR response. The fact
that all the four MR curves can be scaled onto a single curve
suggests there is a universal scattering mechanism. Kohler's
rule\cite{10} suggests that the MR of a material is a universal
function of $\mu$$B$: $R(B)$/$R$($B$=0)=$F$($\mu$$B$). It is
common\cite{10} that at low field when $\mu$B$\ll$1,
$F$($\mu$$B$)$\approx$1+($\mu$$B$)$^2$ , as a result of the
Lorentz force deflection of carriers, with $\mu$ as the carrier
mobility. As shown in Fig2.b, our low $T$ MR data are in good
agreement with Kohler's rule and exhibit a $B^2$ dependence at low
$B$. Using Kohler's rule, one can estimate the mobility $\mu$ from
the parabolic MR at low $B$\cite{18}. We estimate that
$\mu$$\approx$1000, 525, 1500, and 750 cm$^2$/V$\cdot$s for our
nanoribbon sample 1, 2, 3 and 4 at lowest $T$. These mobility
values are somewhat lower than values previously reported on
Bi$_2$Se$_3$ nanoribbons\cite{8,9} and are about 10 times lower
than high quality bulk single crystals\cite{19, 20, 21, 22}.
According to conventional MR theory of metals,
MR$\propto$($\mu$$B$)$^2$ and saturates at high field
($\mu$$B$$>$1). However, our samples did not show saturation of MR
at high $B$. Instead, the MR shows a crossover from $B$$^2$ to
linear $B$ dependence (Fig.2b). Based on the estimated value of
$\mu$, we can see that the crossover from $B^2$ to linear $B$
dependence in our MR data in Fig.2 indeed happens at
$\mu$B$\sim$1. The high field behavior of MR$\propto$$B$ in Fig.2
is reminiscent of Abrikosov's QLMR\cite{11}, which occurs at
$\hbar$$\omega_c$$>$$ E_F$, when all electrons coalesce into the
lowest Landau level. Here $\hbar$ is the reduced Planck's
constant, $\omega_c$=$eB$/m$^*$ is the cyclotron frequency with
$m^*$ being the electron effective mass, $e$ being the electron
charge and $E_F$ being the Fermi energy. Using our estimates of
Fermi momentum and electron
concentration(10$^{17}$-10$^{18}$/cm$^3$) described later in the
paper, the emerging of QLMR at $B>\sim$5T in Fig.1 would
correspond to around 3 Landau levels being filled. It is worth to
note that our low $T$($<$90K) MR data and analysis are in
qualitative agreement with bulk InSb where the QLMR is also found
to emerge at high filling factors\cite{12}. With this
understanding of MR at low $T$, it is enlightening to re-examine
the overall evolution of $R(B)$ from 2K to 300K in Fig.1b. It is
clear that as $T$ rises, the high field linear $B$ behavior
persists into lower field, suggesting that the QLMR is actually
more significant at high temperatures. Since most likely bulk
electrons coexist with surface electrons in the sample, the
disappearance of quadratic MR at high temperature is possibly
related to the different contributions from bulk and surface
states at different $T$.

\begin{figure}
\includegraphics[width=80mm]{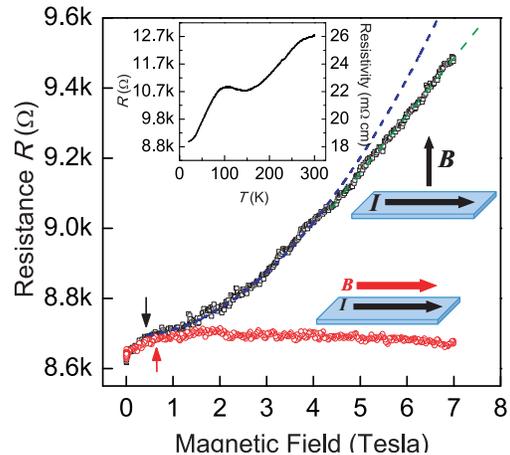}
\caption{\label{Fig.3} (color online) The perpendicular MR
compared with the longitudinal MR at $T$=10K for Bi$_2$Se$_3$
nanoribbon sample $\#$2. The blue dashed line shows the $B^2$
dependence of MR below 4Telsa. The black and red arrows mark the
initial rise of the MR which is attributed to boundary-scattering
effect. The inset shows the resistance and resistivity vs $T$ data
at $B$=0.}
\end{figure}

Now let us turn to the comparison of MR between different field
orientations. Zero field $R$ vs $T$ of nanoribbon sample $\#$2 is
shown in Fig.3 inset. The peak around $T$=100K in the $R(T)$ curve
for this sample resembles those of low doped Bi$_2$Se$_3$ bulk
samples\cite{19}. Assuming homogeneous conduction through the
sample, the 3D resistivity is shown on the right axis of Fig.3
inset using the size of this nanoribbon measured by atomic force
microscopy (AFM) (length $L$=2.34$\mu$m, width $W$=600nm and
thickness $H$=80nm). Fig.3 main panel compares the 10K MR of
sample $\#2$ in perpendicular vs. longitudinal magnetic fields.
The perpendicular MR exhibits a quadratic behavior below 4T and is
much larger than the longitudinal one. It can be noticed that for
both field orientations, there is an initial step-like rise of MR
at low $B$ ($<$1T) as marked by the black and red arrows. This
feature is more salient in samples with lower noise (e.g. sample
$\#$3 in Fig.4 below) and only observed at low $T$ (10K or lower).
We attribute this step-like rise of MR at the lowest $B$ to the
boundary scattering of electrons undergoing cyclotron motion in
our nanoribbons with finite size. A similar effect was observed
for electrons in Bi and Sb nanowires\cite{23,24} as well as 2D
electrons in GaAs heterostructure samples with narrow
width\cite{25}. Briefly, due to the finite size of sample, the
bending of electrons' trajectories by the Lorentz force enhances
the surface scattering when $B$ increases from zero, and such
surface scattering leads to the rise of sample resistance.
However, as $B$ increases to a critical field $B_c$, where the
size of the cyclotron orbital is comparable with the size of
sample, the increase of surface scattering rate slows down and
therefore the resistance rise stops. At $B>B_c$, we do not detect
any features like the A-B oscillations in the longitudinal MR,
presumably due to the lower mobility and shorter coherence length
of our sample than those in Ref.\cite{8}.
\begin{figure}
\includegraphics[width=75mm]{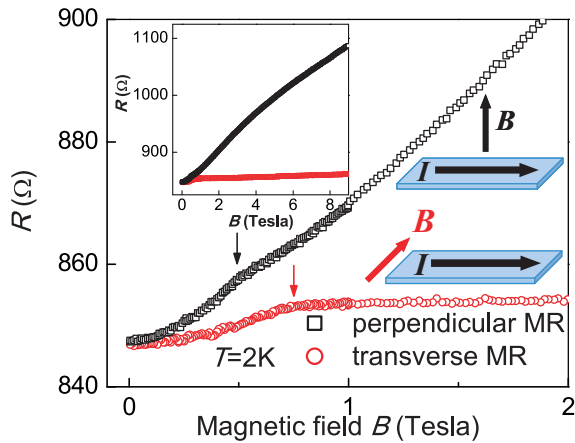}
\caption{\label{Fig.4} (color online) Perpendicular (black
squares) vs transverse (red dots) MR of Bi$_2$Se$_3$ nanoribbon
sample $\#$3 at $T$=2K. The nanoribbon is 900nm wide and 160nm
thick. The black/red arrow highlights the critical magnetic field
$B_c$ for the boundary-scattering (see text). The inset shows the
data on a larger scale ($B$=0-9Tesla).}
\end{figure}

We also compared the perpendicular vs transverse ($B$ parallel to
sample surface but transverse to current) magneto-transport of
Bi$_2$Se$_3$ nanoribbon sample $\#$3. Our experiment shows that
the transverse MR is also negligible except for the initial
step-like increase of MR at low $B$, in a similar way to
longitudinal MR, as shown in Fig.4a and the inset. Due to the low
noise of sample $\#$3, the finite-size-induced boundary scattering
effect in the low $B$ regime of MR can be clearly resolved, with
$B_c$$\sim$0.5T for perpendicular field and $B_c$$\sim$0.75T for
transverse field. The critical field $B_c$ corresponds to the
condition $r_c$ =$d$/2 where $r_c$=$\hbar$$k_F$/$eB$$_c$ is the
cyclotron radius and $d$ is the smallest dimension of the sample
in the plane of cyclotron motion\cite{23,24,25,26}. Since both the
length, width and thickness of our nanoribbons are known (measured
by AFM), we can use the critical field $B_c$ and criterion
$r_c$=$\hbar$$k_F$/$e$$B_c$=$d$/2 to  estimate $k_F$ , the Fermi
momentum, as suggested first by Chambers\cite{26}. For the
perpendicular field case, the relevant $d$ is the sample width
$W$, and we estimate a
$k_F$$\sim$$WeB_c$/2$\hbar$=3.6$\times$10$^8$m$^{-1}$ using
$W$=900nm and $B_c$$\sim$0.5T. For the transverse field
configuration, the relevant $d$ is the sample thickness $H$, and
we estimate a $k_F$$\sim$$HeB_c$/2$\hbar$=1$\times$10$^8$m$^{-1}$
using $H$=160nm and $B_c$$\sim$0.75T. Note that $k_F$ extracted in
the perpendicular field configuration is always a few times larger
than the transverse or longitudinal case for our nanoribbons. This
difference is at odds with the Fermi surface topology of bulk
Bi$_2$Se$_3$ probed by Shubnikov de-Haas effect\cite{21}, which
shows that $k_F$ along c-axis (perpendicular to nanoribbon
surface) should be slighter larger than the in-plane $k_F$ if bulk
electrons are involved. Since the $k_F$ estimated from the
boundary scattering effect is an averaged $k_F$ over the cyclotron
orbit in the plane perpendicular to $B$, this discrepancy suggests
that $k_F$ perpendicular to nanoribbon surface is very small and
is likely due to the confinement effect and/or more dominant
contribution from the surface states in nanoribbon samples.
Nevertheless, the magnitude of $k_F$ is comparable with a 3D
electron concentration on the order of 10$^{17}$-
10$^{18}$/cm$^3$\cite{21} or a 2D electron concentration on the
order of 10$^{12}$/cm$^2$. To make the quasi-2D surface states
dominate over bulk states, it is expected the carrier
concentration should be low. Thus the relatively low carrier
concentration inferred here is reasonable.

Finally, we make a few comments on separating the the surface
states from bulk electrons in the MR effects. While all the
existing data in literature suggest that both the bulk and surface
electrons contribute to the electrical transport, the linear MR we
observed at high $T$ is quite consistent with the topological
surface states with linear dispersion and hard to explain in
conventional theory without introducing strong sample
inhomogeneity. Although the boundary-scattering-induced MR effect
at low field has been discussed in a 3D picture for Bi and Sb
nanowires\cite{23,24}, it does not exclude the possibility of
being originated from surface electrons. What is more important is
that our study of MR in different field orientations reveals that
$k_F$ perpendicular to nanoribbon surface is much smaller than
$k_F$ in the plane. This contradicts the simple 3D Fermi surface
picture\cite{21} but is more consistent with having an origin from
quasi-2D surface electrons. All this suggests that the MR effects
observed are non-trivial and may have connection with topological
surface states with linear dispersion spectrum.

In summary, we report MR of chemically synthesized Bi$_2$Se$_3$
nanoribbons under various magnetic field orientations. When the
magnetic field is parallel to the surface of nanoribbon (a-b
plane), the MR effect is much smaller than in a perpendicular
field. At the lowest $B$($<$1T), we observe boundary scattering
induced MR, from which it is concluded that $k_F$ perpendicular to
surface is much smaller than $k_F$ in the plane. As $B$ increases
such that boundary scattering is unimportant, the low $T$
perpendicular MR exhibits a $B^2$ dependence which crosses over to
a non-saturating linear behavior at high field. This linear MR
extends to zero field as $T$ raises. The linear MR is attributed
to Abrikosov's QLMR and seems to be consistent with the existence
of electrons having linear dispersion spectrum, as predicted in
the topological insulator theory.

X.P.A. Gao acknowledges P.B. Littlewood for discussion, ACS
Petroleum Research Fund (grant 48800-DNI10) and NSF (grant
DMR-0906415) for financial support.


\begin{references}
\bibitem{1}S.C. Zhang, {Physics} {\bf 1}, 6 (2008).
\bibitem{2} J.E. Moore, {Nature} {\bf 464}, 194 (1958).
\bibitem{3} M.Z. Hasan, C.L.Kane, {arXiv:1002.3895v1}.
\bibitem{4}H.J. Zhang, {\it et al}, {Nature Phys.} {\bf 5}, 438 (2009).
\bibitem{5}D. Hsieh,{\it et al},{Nature} {\bf 452}, 970 (2008).
\bibitem{6}Y. Xia, {\it et al}, {Nature Phys.} {\bf 5}, 398 (2009).
\bibitem{7}D. Hsieh,{\it et al}, {Nature} {\bf 460}, 1101 (2009).
\bibitem{8}H.L. Peng, {\it et al}, {Nature Mat.} {\bf 9}, 225 (2010).
\bibitem{9}D. Kong, {\it et al}, {Nano Lett.} {\bf 10}, 329 (2010).
\bibitem{11}A.A. Abrikosov, { Phys. Rev. B} {\bf 58}, 2788 (1998).
\bibitem{carriertype}Tuning sample resistance by a backgate
shows n-type behavior for all our samples, indicating electron as
the majority carrier.
\bibitem{10}J.L. Olsen, {Electron Transport in Metals}, Interscience, New York, (1962).
\bibitem{12}J.S. Hu, T.F. Rosenbaum, {Nature Mat.} {\bf 7}, 697 (2008).
\bibitem{13}M.M. Parish, P.B. Littlewood, {Nature} {\bf 426}, 162 (2003).
\bibitem{14}R. Xu, {\it et al}, {Nature} {\bf 390}, 57 (1997).
\bibitem{16}A.A. Abrikosov, {Europhys. Lett.} {\bf 109},49,789 (2000).
\bibitem{18}G.R. Hyde, H.A. Beale, I.L. Spain, J.A. Woollam, J. Phys. Chem. Solids {\bf 35}, 1719 (1974).
\bibitem{19}J.G. Checkelsky,{\it et al}., {Phys. Rev. Lett.} {\bf 103}, 246601 (2009).
\bibitem{20}J.G. Analytis, {\it et al}., {arXiv:1001.4050}.
\bibitem{21}K. Eto, {\it et al}., {arXiv:1001:5353}.
\bibitem{22}N.P. Butch, {it et al}., {arXiv:1003.2382}.
\bibitem{23}J. Heremans, {\it et al}., {Phys. Rev. B} {\bf 61}, 2921 (2000).
\bibitem{24}J. Heremans, C.M. Thrush, Y.M. Lin, S.B. Cronin, M.S. Dresselhaus, {Phys. Rev. B} {\bf 63}, 085406 (2001).
\bibitem{25}T.J. Thornton, M.L. Roukes, A. Scherer, B.P. Van de Gaag, {Phys. Rev. Lett.} {\bf 63}, 2128 (1989).
\bibitem{26}R.G. Chambers, {Proc. R. Soc. London, Ser.A.} {\bf 202}, 378 (1950).
\end{references}
\end{document}